\documentclass{ws-ijmpa}
\def\uubar{\bar{u}u}

\def\facpi{{1\over 16\pi^2\,}}
\newcommand{\lbl}[1]{\label{eq:#1}}

\newcommand{\be}{\begin{equation}}
\newcommand{\en}{\end{equation}}
\newcommand{\bea}{\begin{eqnarray}}
\newcommand{\ena}{\end{eqnarray}}
\newcommand{\trace}[1]{\langle #1 \rangle}
\newcommand{\Dslash}{\mathrel{%
\setbox0=\hbox{$D$}\copy0\kern-0.8\wd0\hbox{\slash}}}
\newcommand{\lapprox}{\mathrel{%
\setbox0=\hbox{$<$}\raise0.6ex\copy0\kern-\wd0\lower0.65ex\hbox{$\sim$}}}
\newcommand{\gapprox}{\mathrel{%
\setbox0=\hbox{$>$}\raise0.6ex\copy0\kern-\wd0\lower0.65ex\hbox{$\sim$}}}

\begin{document}

\markboth{Bachir Moussallam}{Chiral Effective Action of QCD}

\catchline{}{}{}{}{}

\title{Chiral Effective Action of QCD: Precision tests, Questions and
Electroweak extensions}

\author{Bachir Moussallam}
\address{Institut de Physique Nucl\'eaire, Universit\'e Paris-Sud,
91406 Orsay, France}



\maketitle


\begin{abstract}
This talk first discusses some aspects of the chiral expansion with
three light flavours related to the (non) applicability of the OZI rule.
Next, the extension of ChPT to an effective theory of the full standard model
is considered. Some applications of a systematic description of the 
coupling constants by sum rules (e.g. to the determination of quark masses
and $K_{l3}$ decays) are presented.
\end{abstract}

\section{Introduction}
A major implication of confinement in QCD is to generate an
ordered structure for the ground state\cite{thooft79}.
As a consequence, at low energies, the non-perturbative 
QCD dynamics can be described by an effective theory 
(called ChPT)\cite{weinberg,gl84,gl85}.
In this approach, an expansion in either two or three of the
light quark masses is performed and it is important to verify its accuracy.
This is now becoming possible as many observables have been computed 
at NNLO and impressive  progress has also been achieved on the 
experimental side. 
A summary of the most interesting results concerning the 
$SU(2)$ expansion are presented by Irinel Caprini at this conference. 
The $SU(3)$ chiral expansion is of considerable practical interest 
as it embodies the whole field of kaon physics.
The difference beween the $SU(2)$ and the $SU(3)$ expansions is not simply
that the expansion parameter is larger for $SU(3)$. It reflects also
a difference between the respective chiral vacuums, which is an OZI
suppressed dynamical effect.
This topic is discussed in sec. 2 below.   
The computation of radiative corrections is important for precision tests
and for properly dealing with isospin breaking and this has motivated an 
extension of ChPT. Some aspects and applications 
of this extension are presented in sec.3.
\section{$SU(3)$ chiral expansion and the OZI rule:}
According to the OZI rule (or, equivalently, the large $N_c$ expansion)
the role of the light quarks in the dynamics of chiral symmetry breaking
in QCD is  suppressed as compared to that of the gluons.
Consequently, if one sets 
$N^0_f$ quark masses  equal to zero, the value of 
the chiral condensate $\Sigma$ should be essentially independent of $N^0_f$. 
That this is unlikely to be true is suggested by the Banks-Casher
formula\cite{bankscash} which relates $\Sigma$ to the gluon averaged 
density of small eigenvalues of the Dirac operator.
The dependence upon $N^0_f$ shows up in the integration measure through
a factor $[\det(i\Dslash)]^{N^0_f}$. Obviously, this tends to suppress
the weight of gluon configurations which generate a high density
of small eigenvalues. 
In fact, as we have heard from Th. Appelquist's talk, a critical value
for $N^0_f$ can be argued  to exist  
$N_f^{0\ crit}\simeq 6$ above which spontaneous breaking
of chiral symmetry will no longer occur. One therefore expects
the $SU(3)$ condensate to be reduced
\be\lbl{ineqcond}
\trace{\uubar}_{N^0_f=3} < \trace{\uubar}_{N^0_f=2}
\en

What is the size of this reduction and what 
are the physical implications?
The OZI rule starts to manifests itself in the $SU(3)$ chiral Lagrangian at 
order $p^4$ and implies that 
the coupling constants $L_4$, $L_6$ and the combination 
$L_2-2L_1$ are suppressed\cite{gl85}. How can one determine 
these couplings? It was first noted 
in refs.\cite{kl4rigg,kl4bij} that $L_2-2L_1$ 
can be determined from experimental $Kl_4$ 
decay data. 
Both $L_4$ and $L_6$ can be determined from form-factors associated with the
isospin zero scalar currents $\bar{s}(x) s(x)$ 
and $\bar{u}(x)u(x)+\bar{d}(x)d(x)$. 
These scalar form-factors are not directly accessible to experiment. 
However, it was shown in ref.\cite{dgl}
that they can be determined in the low-energy region $E\lapprox 1$ GeV 
from $\pi\pi\to \pi\pi$ and $\pi\pi\to K\bar{K}$ amplitudes, 
based on analyticity properties and a few plausible assumptions/approximations
(notably a simplified treatment of unitarity in the high energy region 
and a minimal implementation of the Brodsky-Lepage asymptotic 
scaling behaviour\cite{brodskylep} ). 
This construction provides a way of fixing  
the Kaplan-Manohar ambiguity\cite{kaplanman}. 
Table 1 shows the results for  $L_4$ and $L_6$ obtained in this manner
in ref.\cite{bm00} and more recently in ref.\cite{bijdontff} who 
computed the scalar form factors up to $O(p^6)$. Significant violation
of the OZI rule is observed. The value of $L_6$ corresponds to a reduction
of the $SU(3)$ condensate by approximately a factor of two. 
\begin{table}[h]
\tbl{Evaluations of $L_4$, $L_6$ using scalar form factors. }
{\begin{tabular}{|l|l|l|}\hline
                     & $10^3\,L^r_4(m_\rho)$ &  $10^3\,L^r_6(m_\rho)$\\ \hline
OZI rule\cite{gl85}                          & $-0.3\pm 0.3$& $-0.2\pm 0.5$\\
Scalar form-factors, sum-rule\cite{bm00}     & $+0.3\pm 0.2$& $+0.3\pm 0.3$\\
Scalar form-factors, $O(p^6)$\cite{bijdontff}& $+0.4\pm 0.2$& $+0.1\pm 0.3$\\
 \hline
\end{tabular}}
\end{table}

A sensitive probe of the $SU(3)$ chiral expansion is 
the $\pi K$ scattering process.
Based on the $O(p^4)$ calculation of ref.\cite{bkm2}
it was pointed out\cite{abm} that $L_2-2L_1$ as well as $L_4$ can
be determined from $\pi K\to\pi K $ and the crossing symmetric 
amplitude $\pi\pi\to K\bar K$.
On the experimental side, data of good accuracy 
exists in the medium energy region $1\lapprox E\lapprox 2.5$ GeV 
for both $\pi K$ and $\pi\pi\to K\bar{K}$. The general properties of
analyticity, crossing and elastic unitarity allow one to extrapolate these
experimental results down to low energies. More precisely, 
as first shown by Steiner and by Roy\cite{steiner} a set of 
integral equations can be written down for the $l=0$ and $l=1$ 
partial-waves. A set of equations of this type was recently derived 
and analyzed using as input, for the first time, 
the available high-statistics experimental data\cite{dbmpik}. 

\begin{table}[ht]
\tbl{$L_i$ couplings ($\times 10^3$) from $\pi K$ and from $Kl_4$}
{\begin{tabular}{|llll|l|}\hline
$\phantom{1.3}L_2$ & $\phantom{-4.5}L_3$ & ${\phantom{-}L_2-2L_1}$ 
& {$\phantom{0.5}L_4$} & $\times 10^3$ \\ \hline
$1.3\pm0.1$ & $-4.5\pm0.1$ & $-0.8\pm0.2$ & $0.5\pm 0.4$ & $\pi K$ \\ 
$1.5\pm0.2$ & $-3.2\pm0.8$ & $+0.6\pm0.5$ & $\phantom{0.5}-$ & $Kl_4$ \\ \hline
\end{tabular}}
\end{table}
Table 2 shows the results for the $O(p^4)$ coupling constants 
$L_1$, $L_2$, $L_3$, $L_4$ determined
from the $\pi K$ amplitude as obtained from the  Roy-Steiner solution.
For the coupling $L_4$, one observes 
good agreement with the determination from the scalar form-factors.
The table also shows the determination 
(from ref.\cite{Kl4bijtal})
which uses data on $K_{l4}$ decays. The results for $L_1$ and $L_3$ 
are in reasonably good agreement. The only discrepancy concerns the OZI
suppressed combination $L_2-2L_1$. This suggests that the $SU(3)$ 
chiral expansion is functioning properly and that including the $O(p^6)$
corrections should resolve the discrepancies.

The computation of the $\pi K$ amplitude at order $p^6$ has recently been 
completed\cite{bijdontpik}. The expression involves 
28 couplings $C_i$ from the $O(p^6)$ Lagrangian\cite{cgep6}. 
In order to make predictions, 
one may    estimate the relevant $C_i$'s from
resonance saturation models. This approach has proved reasonably
successful for the $O(p^4)$ couplings $L_i$\cite{egpr}.
Clearly, the task is harder in the case of the $C_i$'s 
because many more terms in the resonance Lagrangians must be considered.
The numerical predictions presented in ref.\cite{bijdontpik} are based on a
rather minimal resonance model. Their results  concerning the S-wave
scattering lengths agree rather well with those generated
by solving the Roy-Steiner equations (see Table 3 below).
\begin{table}[h]
\tbl{$S$-wave $\pi K$ scattering lengths }
{\begin{tabular}{|lll|}\hline
\             & $m_\pi\,a_0^{1\over2}$ & $m_\pi\,a_0^{3\over2}$\\ \hline
ChPT $O(p^6)$ & $0.212$                & $-0.051$ \\
Roy-Steiner   & $0.224\pm 0.022$       & $-0.045\pm 0.008$  \\
\hline
\end{tabular}}
\end{table}
However, a more detailed comparison of the chiral and the dispersive
amplitudes shows several discrepancies and, sometimes,  
unphysically large $O(p^6)$ contributions. 
Efforts to refine the estimates of the  $C_i$'s are now needed  
in order to improve the predictivity of the $SU(3)$
expansion and our understanding of its workings.

\section{Electroweak extensions}
The ChPT formalism is well adapted to the computation of radiative 
corrections. The extension, which consists in treating
the photon as a dynamical field was first developed by Urech\cite{urech95}. 
A further extension has been performed, 
allowing the light leptons to be treated
dynamically as well\cite{knechtlept} (which is necessary for 
computing radiative corrections in semi-leptonic
processes). A natural chiral counting for the charge
is to set $O(e)\sim O(p)$ and, for a lepton field $l$, to set 
$O(l)=O(p^{1/2})$. 
The resulting setup represents the
low-energy effective theory of the full standard model since it includes
all of the light particles. 

The extended chiral Lagrangian involves a number of new coupling
constants: one coupling $C$ at order $e^2$, then 13
couplings\cite{urech95} $K_i$ at order $e^2 p^2$ and 7 
couplings\cite{knechtlept} $X_i$
at order $e^2 l^2 p$. A basic property of the chiral
couplings in this sector is that they can be expressed as
sum rules involving a QCD Green's function and the photon 
propagator.  This was noted long ago\cite{dgmly}
for the $\pi^+-\pi^0$ mass difference. Systematic 
generalizations have been derived
for the couplings $K_i$ which involve 2,3 and 4-point 
QCD Green's functions\cite{bm97}. These sum rules can be used
to estimate the coupling constants based on light resonance
models. It is important to
constrain the models such as to reproduce 
the proper QCD short distance behaviour of the relevant 
Green's functions. 
Several sum rules actually diverge in four dimensions. 
The divergences are canceled by direct contributions from 
QED+QCD counterterms. 
As a consequence, the 4 couplings
$K_9$, $K_{10}$, $K_{11}$ and $K_{12}$ depend not only on the chiral scale
$\mu$ but also on the short distance renormalization scale $\mu_0$.
\subsection{``Strong'' quark mass}
As an application, let us consider the 
definition of a ``strong'' quark mass $\bar m_f$, i.e. a quantity which runs
according to QCD only and not QCD+QED as the physical mass $m_f$ does. 
This issue was recently 
discussed and illustrated in several models\cite{rusetsky}. 
ChPT provides a very simple answer to this question
\be
\bar m_f(\mu,\mu_0) = m_f (\mu_0) ( 1+ 4 e^2 Q_f^2 [K_9^r(\mu,\mu_0)
+ K_{10}^r(\mu,\mu_0)] ) +O(e^4)
\en
The strong mass $\bar m_f$ depends on two scales
in accordance with ref.\cite{rusetsky}. 
In practice, the quark mass ratios that can be extracted from low energy data
using ChPT concern precisely these ``strong'' masses since all the ratios
are QCD renormalization group 
invariants (up to $O(e^4)$ ). The physical mass ratios can then be determined
knowing the value of $K_9+K_{10}$. This combination satisfies the following
sum rule,
\bea
&& K_9^r + K_{10}^r ={1\over 8F_0^2(m_K^2-m_\pi^2)}
\int d^d x 
\trace{0\vert T[ V^{ud}_\mu(x) V^{du}_\nu(x)
- V^{us}_\mu(x) V^{su}_\nu(x)] \vert0} D^{\mu\nu}_\gamma(x)
\nonumber\\
&&\phantom{K_9^r + K_{10}^r  }
+{3C\over64\pi^2 F_0^4}\left[
{m_\pi^2\log{m_\pi^2\over m_K^2}\over  m_\pi^2
-m_K^2}+\log{m_K^2\over\mu^2}\right]
+(Z_s -Z_2)  + O(m_s )
\nonumber
\ena
where $Z_2$ and $Z_s$ are QED renormalization parameters.
The integrand can be rewritten in terms of spectral functions which,
in principle, can be obtained from data on $\tau$ decays into hadrons
with $S=0$ and $S=1$. At present, the $\tau$ decay data with $S=1$ 
has not enough statistics for a fully quantitative evaluation 
but this will become possible with the advent of $\tau$-charm factories.
\subsection{$X_i$ sum rules}
One might anticipate that the chiral couplings $X_i$, which appear 
at $O(p^4)$ in the chiral Lagrangian with dynamical leptons and photons 
should also satisfy sum rules. 
These have been investigated recently\cite{xidescotes}. A basic ingredient 
is the  calculation\cite{bratli}, at order one
loop in the standard model, of the semi-leptonic decay amplitude 
$l(p)\to \bar{u}(q)+d(q')+\nu(p')$.
The matching to the analogous calculation performed with  ChPT
must be done in two steps. One first considers a four-fermion
Fermi type effective theory ${\cal L}_{Fermi}$
obtained by integrating out the heavy bosons
in the SM. This effective theory is valid in a range of energies $E<< M_W$
and $E\gapprox 2$ GeV such that it makes sense to treat the quarks and gluons
perturbatively. Matching the calculation at order one loop in this
theory and in the SM determines a set of four counterterms to be added to
${\cal L}_{Fermi}$. From a technical point of view it is wise to
rely on Pauli-Villars rather than dimensional regularization which
avoids all problems with $\gamma^5$. In a second step one performs
a matching with ChPT. The objects that one matches 
are spurion-Green's functions which are obtained by performing functional
differentiations of the generating
functional with respect to the charge spurions.  
This method generates in a straightforward way 
the sum rules for the $X_i$ parameters. 
As an illustration the expression for the combination
$X_6-4 K_{12}$, which corresponds to 
Sirlin's\cite{sirlin} logarithmically enhanced universal factor $S_{EW}$ is:
\be
X_6^r(\mu)-4K^r_{12}(\mu)\simeq {1\over32\pi^2}\int_0^{M^2_Z}dx\,[
\Gamma_{VV}(-x)+\Gamma_{AA}(-x)]
+\facpi \left[ -6\log{M_Z\over\mu} +{5\over2}\right]
\en
where $\Gamma_{VV}$, $\Gamma_{AA}$ are the form-factors associated with
the matrix elements $\trace{0\vert V_\mu V_\nu \vert\pi}$ and 
$\trace{0\vert A_\mu A_\nu \vert\pi}$. 
The sum rule allows to estimate the contributions from
the resonance region and the perturbative $\alpha_s$ one in addition 
to the large logarithm.

\begin{table}[b]
\tbl{Isospin breaking $K_{l3}$ form-factor ratio and mass ratio}
{\begin{tabular}{|ccccc|}\hline
\ & Dashen & EM sum rules & MILC\cite{milc} & experiment \\ \hline
$R$ & 41.5  & 30.5  & 33.3 & $-$ \\
$r_{+0}$                    & 1.020 & 1.030 & $-$  & $1.040\pm 0.010$ \\ \hline
\end{tabular}}
\end{table}
Several  new measurements of $K^0\to \pi^+ l^- \nu$ 
decays\cite{K0exps} as well as  $K^+\to \pi^0 l^+ \nu$ decays\cite{K+exps} 
have been performed with the aim of refining the determination of $V_{us}$. 
The compatibility of the two sets of results was questioned\cite{cirig04}. 
As a measure of this, let us consider the ratio 
of the $f_+$ form factors which, in ChPT, reads
\be
r_{+0}\equiv 
{f^{K^+\pi^0}_+(0)\over f^{K^0\pi^+}_+(0)}= 1+{3\over4}\,{1 \over
R } + O(p^4,e^2 p^2),\quad R={ 2m_s -m_u-m_d\over 2(m_d-m_u) } 
\en
Complete expressions for the $O(p^4)$ contributions (including 
electromagnetic ones) can be found in ref.\cite{cirig04}. The 
sum rules allow one to estimate all the $K_i$ and $X_i$ 
coupling constants involved. 
The quark mass combination $R$ can be determined from
the $K^+-K^0$ mass difference as a function of the mass ratio\cite{gl85} 
$r=2m_s/ (m_u+m_d)$ (for which we will use $r=27.1$ as obtained by 
MILC\cite{milc}). 
Again here, the electromagnetic contributions
beyond the leading order result (given by Dashen's theorem\cite{dashen})
can be determined from the sum rules. Numerical results are collected in
table 4.
The table shows that the isospin breaking ratio $R$ is likely to be 
significantly different from its leading order determination. An important test
of $SU(3)$ ChPT will be to verify the compatibility
of the value of $R$ with the $\eta\to3\pi$ process\cite{leutw96}.

\section*{Acknowledgements}
Partial support by EU RTN contract HPRN-CT-2002-00311 (EURIDICE) and 
FP6 programme (HadronPhysics) contract RII3-CT-2004-506078 are aknowledged.

\end{document}